\documentclass[multphys,vecphys]{svmult}

\usepackage{makeidx}     
\usepackage{graphicx}    
\usepackage{multicol}    

\makeindex             

\def\p{$e^\pm \;$}

\begin{document}

\title*{On the Central Engine of short Gamma-ray Bursts}
\author{Stephan Rosswog\inst{1}\and
Enrico Ramirez-Ruiz\inst{2}}
\institute{Dep. Physics \& Astronomy, University of Leicester, Leicester LE1 7RH, UK
\texttt{sro@astro.le.ac.uk}
\and Institute of Astronomy, Cambridge, CB3 OHA, UK
\texttt{enrico@ast.cam.ac.uk}}
%
%
\maketitle

\section{Introduction}

While there is so far no direct evidence linking coalescences of double neutron 
star systems to gamma-ray bursts (GRBs), there can be no doubt about the plausibility
of this system, at least for the subclass of short GRBs. Neutron star binaries can 
provide huge reservoirs of gravitational binding energy and are expected to lead 
naturally to short overall durations with variations on millisecond time scales. 
In the following we want to assess two popular mechanisms to launch a GRB.

\section{Numerical method and simulations}
\label{sec:2}
\begin{figure}
\centering
\includegraphics[height=6cm]{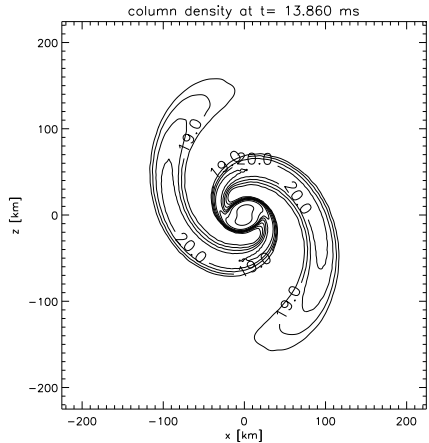}
\caption{Column density of an initially corotating neutron star binary system.
For this simulation more than $10^{6}$ SPH particles have been used.}
\label{cont}    
\end{figure}

We have performed global, 3D simulations of the last stages prior to
the coalescence and followed the subsequent hydrodynamical evolution for 
about 15 ms. We use a temperature and composition dependent nuclear equation
of state that covers the whole relevant parameter space in density, temperature
and composition \cite{shen98,rosswog02a}. In addition, a detailed,
multi-flavour neutrino treatment has been applied to account for energy losses
and compositional changes due to neutrino processes. The neutrino treatment
and the results concerning the neutrino emission have been described in detail
in \cite{rosswog03a}.
To solve the hydrodynamic equations we use the smoothed particle hydrodynamics 
method (SPH), the simulations are performed with up to more 
than a million SPH particles, see Figure \ref{cont}. The details of the production runs as well as 
those of several test runs can be found in 
\cite{rosswog02a,rosswog03a,rosswog03b}. Results focusing particularly on gamma-ray bursts
have been presented in \cite{rosswog02b,rosswog03b,rosswog03c}.

\section{Assessment of the GRB mechanisms}

The energy released in a GRB  represents just a minor fraction of the released 
gravitational binding energy (several times $10^{53}$ erg). This allows
(in principle) for a plethora of possible burst mechanisms. The most 
popular of these are neutrino annihilation 
\cite{jaroszinski93,mochkovitch93,ruffert97,popham99,ruffert99,asano00,salmonson01} 
and the extraction of rotational energy via magnetic fields
\cite{narayan92,thompson93,usov92,mesz97,katz97,piran99,mesz02}.

\subsection{Neutrino Annihilation}

\begin{figure}
\centering
\includegraphics[height=8cm,angle=90]{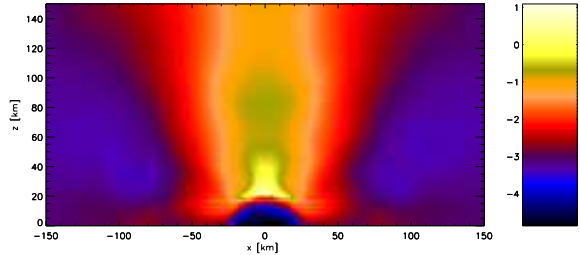}
\caption{Relativistic jet launched from a merger remnant via $\nu \bar{\nu}$ annihilation 
(upper half-plane; for details see \cite{rosswog03c}). Colour coded is the logarithm
of the attainable Lorentz factor.}
\label{jet}       
\end{figure}

\begin{figure}
\centering
\includegraphics[height=6cm]{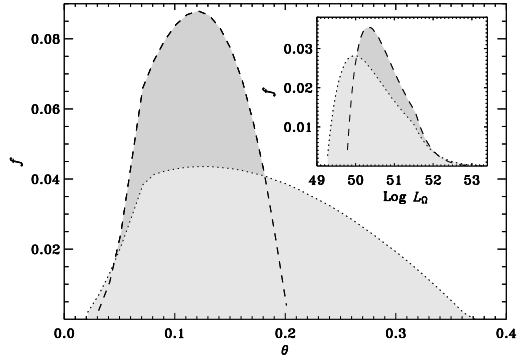}
\caption{Distribution of opening angles and apparent luminosities (inset)
for GRBs launched via $\nu \bar{\nu}$ annihilation and collimated by means 
of a baryonic wind. (The dashed lines refer to $\alpha=3.2$, the dotted to 
$\alpha=3.4$; see text).}
\label{dist}  
\end{figure}

We find the merger remnant to emit neutrinos at a total luminosity of 
$\sim 2 \cdot 10^{53}$ erg/s. The luminosities are dominated by
electron anti-neutrinos, followed by electron neutrinos and -slightly 
less luminous- the heavy lepton neutrinos. Their rms energies are around 15, 8
and 20 MeV, respectively. The bulk of the neutrino emission comes from the 
inner regions of the hot torus that has formed around the central object
of the merger remnant (see \cite{rosswog03a} for details). In this context
we consider two phenomena: (i) neutrinos and anti-neutrinos that annihilate
above the merger remnant into \p pairs and (ii) like in the case of a 
newborn proto neutron star the neutrinos will blow off a strong baryonic 
wind from the remnant.\\
The thick disk with its geometry that is favourable for head-on neutrino 
collisions and its low baryon density along the original binary rotation axis
launches a pair of relativistic jets \cite{rosswog02b}. An example of such a 
jet is shown in Figure \ref{jet}. 
The typical energy provided in this way ($\sim 10^{48}$ erg), 
however, is way below the isotropic energy estimates for short GRBs 
($\sim 10^{51}$ erg) at a redshift of $z=1$ \cite{panaitescu01,lazzati01} and
therefore neutrino annihilation can only be a viable GRB mechanism if it 
goes along with a substantial collimation of the resulting outflow. \\
Such a collimation can be obtained via the ram pressure of the neutrino-driven,
baryonic wind. This hydrodynamic collimation mechanism has been suggested by 
Levinson and Eichler \cite{levinson00}. They find that the jet half opening angle 
at large distances from the source, $\theta$, is determined by the ratio of jet 
and wind luminosity, $\theta \propto L_j/L_w$. Since the luminosity in the wind 
exceeds that of the jet by far, the outflow is collimated into a narrow solid angle. 
Using the theoretical neutron star mass distribution of \cite{fryer01}, calculating
the neutrino emission as a function of the binary system mass and parameterising the 
dependence of the wind luminosity as a function of the neutrino luminosity,
$L_w \propto L_{\nu}^\alpha$, $\alpha=3.2 ... 3.4$ \cite{thompson01}, we find
the distributions of opening angles and luminosities shown in Figure 
\ref{dist}. The broad distribution of opening angles is centred around 
6 degrees, the luminosities around a few times $10^{50}$ ergs/s. This is 
compatible with both the observational constraints on the luminosities
of short GRBs and the estimated neutron star merger rates \cite{rosswog03b}.

\subsection{Magnetic Processes}

The violent fluid motion within the merger remnant will almost certainly amplify
the initial neutron star seed fields tremendously. To date, we are still lacking 
detailed MHD calculations of a neutron star coalescence, so we still have to wave 
our hands and draw conclusions from either simplified analytical models or purely 
hydrodynamic calculations.\\
The equipartition field strength, $B^{\rm{eq}}= \sqrt{8 \pi \rho c_s^2}$ in the central
object of the merger remnant is (depending on the exact position) between $10^{16}$ G 
and $10^{18}$ G, and  $10^{14}$ to $10^{16}$ G in the surrounding torus \cite{rosswog03c}. 
If just the relatively slow wrapping of the field lines via differential rotation 
(and no feedback onto the fluid) is assumed, equipartition will be
reached in the central object within a few tens of seconds (provided it remains stable 
for long enough) and in the torus in around 4 s. Other field amplification  mechanisms
are expected to amplify the field exponentially leading to {\em much} shorter time scales.
The fluid motion within the central object exhibits ``convective cells'' with sizes of $\sim$ 
1 km and velocities of $\sim 10^8$ cm/s. Moreover, the neutrino emission will, like in a
proto neutron star, establish a negative entropy and lepton number gradient and therefore 
drive vigorous convection (e.g. \cite{epstein79}). We find Rossby numbers (ratio of rotational 
and convective time scales), $Ro \equiv T_{\rm rot}/\tau_{\rm conv}$, substantially below 
unity (down to 0.1) and therefore expect that the system can sustain a large scale dynamo. 
Such a dynamo will increase the magnetic field exponentially with an e-folding time close to the
convective time scale. Using the numbers determined from our simulations we find that
equipartition will be reached withing tens of milliseconds. The kinetic energy of
the central object, $E_{\rm kin}$, is large enough for an average 
field strength $\langle B \rangle_{\rm co}= \sqrt{3 \cdot E_{\rm kin}/ R_{\rm co}^3}
\approx 3\cdot 10^{17}$ G, where $ R_{\rm co}$ is the radius of the central 
object. With this average field strength the spin-down time scale is
$\tau_{\rm sd}= E_{\rm kin}/L_{\rm md} \approx 0.2$ s, where $L_{\rm md}$ is the magnetic
dipole luminosity, i.e. $\tau_{\rm sd}$ is of the order of the typical duration of a short GRB.\\
A discussion of further possible magnetic mechanisms can be found in \cite{rosswog03c}. 
All of these mechanisms yield luminosities in excess of a few times $10^{52}$ ergs/s and 
therefore yield typical time scales of order 1 s.

\section{Conclusions}

We have assessed the two most popular mechanisms to produce short GRBs from the coalescence
of neutron star binaries, namely the annihilation of neutrino anti-neutrino pairs and 
magnetic energy extraction mechanisms. We find that $\nu \bar{\nu}$ annihilation provides 
the driving stresses
to launch a pair of relativistic, bipolar jets. To explain the expected isotropized energies
the jets have to be narrowly collimated. This collimation can be provided by the energetic,
neutrino-driven baryonic outflow that goes along with the coalescence. Due to the sensitivity
to the neutrino luminosities even a narrow mass spectrum results in a broad distribution
of opening angles and apparent GRB luminosities.\\
We further expect the initial neutron star magnetic fields to be amplified to values close
to equipartition within fractions of a second. The estimates of the various investigated
magnetic mechanisms all yield very large luminosities that do not require any beaming.
If the current picture turns out to be correct, short GRBs should be composed of 
two distinct components: the one resulting from neutrino annihilation plus wind collimation 
is narrowly beamed and the other, much more energetic, and possibly uncollimated component 
results from the plethora of magnetic mechanisms. 

%
%

%
%



\printindex
\end{document}